\documentclass[aps,prd,onecolumn,groupedaddress,showpacs,nofootinbib,amssymb
]{revtex4}
\usepackage[dvips]{graphicx}
\usepackage{amssymb}
\usepackage{amsmath}
\usepackage{graphicx}
\usepackage{amsfonts}
\usepackage{bm}

\begin{document}

\title{Singular $F(R)$ Cosmology Unifying Early and Late-time Acceleration with Matter and Radiation Domination Era}
\author{S.~D.~Odintsov$^{1,2,3}$}
\email{odintsov@ieec.uab.es}
\affiliation{$^{1)}$Institut de Ciencies de lEspai (IEEC-CSIC),
Campus UAB, Carrer de Can Magrans, s/n\\
08193 Cerdanyola del Valles, Barcelona, Spain}
\affiliation{$^{2)}$ICREA, Passeig LluA­s Companys, 23,
08010 Barcelona, Spain}
\affiliation{ $^{3)}$Tomsk State Pedagogical University, 634061 Tomsk, Russia}
\author{V.~K.~Oikonomou$^{4,5}$}
\email{v.k.oikonomou1979@gmail.com, voiko@sch.gr}
\affiliation{ $^{4)}$Laboratory for Theoretical Cosmology, Tomsk State University of Control Systems
and Radioelectronics (TUSUR), 634050 Tomsk, Russia}
\affiliation{ $^{5)}$National Research Tomsk State University, 634050 Tomsk,
Russia}

\begin{abstract}
We present some cosmological models which unify the late and early-time acceleration eras with the radiation and the matter domination era, and we realize the cosmological models by using the theoretical framework of $F(R)$ gravity. Particularly, the first model unifies the late and early-time acceleration with the matter domination era, and the second model unifies all the evolution eras of our Universe. The two models are described in the same way at early and late times, and only the intermediate stages of the evolution have some differences. Each cosmological model contains two Type IV singularities which are chosen to occur one at the end of the inflationary era and one at the end of the matter domination era. The cosmological models at early times are approximately identical to the $R^2$ inflation model, so these describe a slow-roll inflationary era which ends when the slow-roll parameters become of order one. The inflationary era is followed by the radiation era and after that the matter domination era follows, which lasts until the second Type IV singularity, and then the late-time acceleration era follows. The models have two appealing features: firstly they produce a nearly scale invariant power spectrum of primordial curvature perturbations and a scalar-to-tensor ratio which are compatible with the most recent observational data and secondly, it seems that the deceleration acceleration transition is crucially affected by the presence of the second Type IV singularity which occurs at the end of the matter domination era. As we demonstrate, the Hubble horizon at early times shrinks, as expected for an initially accelerating Universe, then during the matter domination era, it expands and finally after the Type IV singularity, the Hubble horizon starts to shrink again, during the late-time acceleration era. Intriguingly enough, the deceleration-acceleration transition, occurs after the second Type IV singularity. In addition, we investigate which  $F(R)$ gravity can successfully realize each of the four cosmological epochs.
\end{abstract}


\maketitle



\def\pp{{\, \mid \hskip -1.5mm =}}
\def\cL{\mathcal{L}}
\def\be{\begin{equation}}
\def\ee{\end{equation}}
\def\bea{\begin{eqnarray}}
\def\eea{\end{eqnarray}}
\def\tr{\mathrm{tr}\, }
\def\nn{\nonumber \\}
\def\e{\mathrm{e}}

\section{Introduction}

Finite time singularities are timelike singularities that frequently occur in cosmological scenarios of modified gravity. These were firstly classified in Ref. \cite{Nojiri:2005sx}, and this classification uses the physical quantities that can be defined on a three dimensional spacelike hypersurface determined by the time instance that the singularity occurs, which are the energy density, the scale factor and the pressure density. The less severe from a phenomenological point of view is the so-called Type IV singularity \cite{Nojiri:2005sx}, which in addition to the three aforementioned physical quantities, also uses the Hubble rate as a classification criterion, and particularly for this singularity the higher derivatives of the Hubble rate diverge. This form of singularity, belongs to a wider class of singularities, which are soft singularities like sudden singularities, which have been firstly studied in \cite{barrownew}. The interesting features of this singularities which was revealed in Ref. \cite{barrownew}, was that closed universes which obey the energy condition $\rho + 3p>0$, and also $\rho>0$, need not to recollapse, owing to the fact that a pressure singularity could occur before the time instance these universes acquire an expansion maximum. Later on, sudden singularities were concretely studied in \cite{Barrow:2004xh1}, and further developed in \cite{barrow1,barrow2,barrow3,barrow4,barrow5,barrow6,barrow7,barrow8,szy}. Recently, the phenomenological implications of Type IV singularities were thoroughly investigated in Refs. \cite{Barrow:2015ora,noo1,noo3,noo4,noo5}. As was explicitly demonstrated in these works, the Type IV singularity has no effect on the physical quantities of the cosmological system, but these critically affect the dynamical evolution of the cosmological system, and in some cases, even the graceful exit from inflation can be triggered \cite{noo4}. Therefore, these singularities are ``harmless'' for phenomenology and in principle no constraints can be imposed to them since the Universe passes smoothly through these singularities, except for the case of gravitational baryogenesis, which if it occurs, can constrain the Type IV singularities, as was demonstrated in \cite{noo5}. 
  
In this paper we shall present a cosmological model that contains two Type IV singularities and unifies all the cosmological eras of our Universe, the early and late-time acceleration with the radiation and matter domination era. Note that unification of early-time acceleration with dark energy in the context of $F(R)$ gravity was proposed in Ref. \cite{sergnoj}, and was further extended to include the matter dominance era in Ref. \cite{Nojiri:2006gh}. The two Type IV singularities are chosen to occur one at the end of the inflationary era and one at the end of the matter domination era and a similar to this model was introduced in \cite{noo5}, where some phenomenological aspects in the context of gravitational baryogenesis were studied. With regards to the chronology of the cosmological evolution, we assume that firstly our Universe starts its evolution with an inflationary era, preceding the radiation and matter domination eras. The cosmological model is described by a slow-roll inflationary era, which generates a quasi de Sitter evolution, approximately identical to the $R^2$ inflation model \cite{starobinsky}, and for the vacuum $F(R)$ case we study here, this era is followed by the radiation and matter domination eras. We need to stress the fact that for both the radiation and matter domination eras, the effective equation of state (EoS) of the cosmological model is approximately $w\simeq 1/3$ and $w\simeq 0$ respectively, but not exactly equal to $w= 1/3$ and $w= 0$. This is very important for the case of the radiation domination era, because if the EoS parameter was exactly equal to $w=1/3$, the corresponding Ricci scalar for a flat Universe would be zero (since the Hubble rate would be equal to $H(t)=1/2t$), and therefore it would not be possible to provide a vacuum $F(R)$ gravity description without adding extra matter fluids. But in our case, since $w\simeq 1/3$, this fact enables us to provide an approximate vacuum $F(R)$ gravity description at leading order. Near the end of the matter domination era, a second Type IV singularity occurs, and after that, the Universe continues its evolution in an accelerating way. Note that we constrain the form of the Type IV singularities in order to avoid unwanted instabilities in the slow-roll indices. We use the $F(R)$ gravity theoretical framework (for reviews see \cite{reviews1}) in order to realize the cosmological model under study, and we investigate which vacuum $F(R)$ gravity can generate each one of the four aforementioned cosmological eras. In addition, we examine the behavior of the EoS, and as we demonstrate, at early times and late times, the EoS is identical to a nearly de Sitter EoS, while during the matter domination era, it is approximately equal to zero, thus mimicking a matter dominated era. Moreover, the EoS during the radiation domination era is approximately equal to $w\simeq 1/3$, as expected for a radiation domination era. For similar works in the context of modified gravity for which unification of early and late-time acceleration with matter dominated era occurs, see \cite{sergnoj,Nojiri:2006gh,capp,refcq1,refcq2}.

In order to make the study more easy, we first study a preliminary unification model, for which the late and early-time acceleration eras are described in a unified way with the matter domination era. This preliminary model has the same qualitative features with the model that unifies all cosmological eras, but it is more simple so it will make the understanding of the qualitative implications of the two Type IV singularities more easy.

Another important study we perform is related to the evolution of the Hubble radius, or equivalently of the Hubble horizon. The evolution of the Hubble horizon for both the cosmological models we study, at early time behaves as the $R^2$ model of inflation at early times, so the Hubble horizon shrinks in a nearly exponential way, as in most inflationary models, for reviews see \cite{inflation}. After the inflationary era, and during the radiation and matter dominated eras, the Hubble horizon expands, because the Universe is expanding in a decelerating way. Interestingly enough, near the second Type IV singularity at the end of the matter dominated era, the Hubble radius starts to shrink again, since the Universe starts to expand in an accelerating way. The interesting feature is that the deceleration-acceleration transition occurs very close to the second Type IV singularity, and the appearance of the singularity seems to have some effect on this deceleration-acceleration transition. The behavior of the Hubble radius implies that for both the cosmological models we present, the cosmological perturbations relevant for the present day observations, originate from the early-time era, during inflation, and when inflation begun, these were well inside the Hubble radius. Then as inflation proceeded, these perturbations exited the horizon and ``froze'', as in the standard $R^2$ inflation case. During the matter domination era, the Hubble horizon expands, and in the process, the primordial perturbation modes reenter the horizon, and these modes are exactly the ones relevant for observations at present time. We calculate the observational indices for the cosmological models and as we demonstrate these are compatible with both the latest Planck data \cite{planck} but also with the recent BICEP2/Keck-Array data \cite{BICEP2}. Finally, the graceful exit from inflation for both our models occurs when the slow-roll expansion \cite{barrowslowroll} breaks down, which occurs when the slow-roll indices become of the order one.

We need to note that the model at early times chronologically describes the Universe from the slow-roll inflation period and after that, so the period before the slow-roll inflationary era does not interest us (this issue was discussed in \cite{reviews1} and also in Ref. \cite{sergbam08}, where it is shown that realistic $F(R)$ gravity models can cancel the initial singularity too). During the preinflationary period the Universe is assumed to expand in an accelerating way, until it reaches the slow-roll inflationary era, from which point the Universe is described from our model. In the discussion session at the end of the paper, we briefly discuss some interesting preinflationary scenarios that appear in the literature \cite{piao1,piao2,piao3,piao4,wette}.

This paper is organized as follows: In section II we present in brief the details of the background geometry and also some basic properties of the finite time singularities. In section III, we present and describe in detail the essential features of a preliminary unification cosmological model, which will make the understanding of the qualitative features more easy. We also present the singularity structure of the preliminary cosmological model, for the various values of the free parameters of the theory and we specify the values of the free parameters of the model. In addition, we investigate the EoS behavior as a function of the cosmic time, focusing on the early and late-time acceleration eras and also to the matter domination era. In section IV we study the evolution of the Hubble radius and we analyze in detail its behavior as a function of the cosmic time. In section V, by using well known reconstruction techniques, we investigate which vacuum $F(R)$ gravity can realize each evolution era, that is, early and late-time but also the matter domination era. Also, we calculate the spectral index of the primordial curvature perturbations and also the scalar-to-tensor ratio, and we demonstrate that the observational indices are compatible to the latest observational data. In section VI, we present some possible modification of the preliminary unification model but more importantly we present a variant form which can describe all the evolution eras of our Universe in a unified way, namely, the late and early-time acceleration eras with the matter and radiation domination eras. As we demonstrate, the qualitative behavior of the unification model is identical to the preliminary model, so all our previous findings apply to this case too. In addition, we investigate which vacuum $F(R)$ gravity can realize the nearly radiation domination era and also discuss in some detail the qualitative feature of the unification model. A discussion on our results and the concluding remarks follow in the end of the paper.

\section{Essentials of Finite-time Singularities and Brief Description of the Background Geometry}

Before getting to the details of the singular models, for the readers convenience and in order to render the article self-contained, we briefly present some essential information for the timelike singularities that occur in cosmology and we also describe the geometric conventions for the spacetime that we use throughout the paper. We start with the geometry, which is described by the flat Friedmann-Robertson-Walker (FRW) spacetime, the line element of which is equal to,
\be
\label{metricfrw} ds^2 = - dt^2 + a(t)^2 \sum_{i=1,2,3}
\left(dx^i\right)^2\, ,
\ee
with $a(t)$ being the scale factor. Moreover, we assume that the connection is a symmetric, metric compatible and torsion-less, affine connection the Levi-Civita connection. For the flat FRW metric of Eq. (\ref{metricfrw}), the Ricci scalar takes the following form, 
\begin{equation}
\label{ricciscal}
R=6\left(2H(t)^2+\dot{H}(t)\right)\, ,
\end{equation}
with $H(t)$ being as usual the Hubble rate, $H(t)=\dot{a}/a$. 

In cosmology, there are four types of timelike singularities, the Big Rip, the Type II, III and IV singularities, which we now describe in brief. The classification of singularities which we now present, uses the scale factor, the energy density, the effective pressure and the higher derivatives of the Hubble rate, as criteria in order to classify the various singularities. This type of classification was firstly presented in Ref.~\cite{Nojiri:2005sx}, according to which, there are four types of timelike singularities in cosmology, which appear below: 
\begin{itemize}
\item Type I (``Big Rip Singularity''): Among all the finite-time singularities, this is the most ``severe'' from a phenomenological point of view, since all the physical quantities that can be defined on the three dimensional spacelike hypersurface determined by the time instance that the singularity occurs, strongly diverge. Indeed, for a Big Rip singularity \cite{br}, the scale factor, the energy density and the effective pressure diverge.
\item Type II (``The Sudden Singularity''): This singularity type was developed in \cite{Barrow:2004xh1,barrow4}. For this timelike singularity, the scale factor and the energy density remain finite, while the effective pressure diverges, so it is phenomenologically less severe compared to the Big Rip.  
\item Type III: This singularity comes after the Big Rip, since in this case only the scale factor is finite, while the pressure and the energy density strongly diverge at the singularity.
\item Type IV: The Type IV singularity is the less harmful singularity, and it is quite interesting from a phenomenological point of view, as was demonstrated in \cite{noo4}. In this case, all the physical quantities defined on the spacelike hypersurface determined by the time instance that the singularity occurs, are all finite, and only the higher derivatives of the Hubble rate diverge. Particularly, the scale factor, the energy density and the effective pressure are finite, while only the higher derivatives of the Hubble rate diverge, that is, $\frac{\mathrm{d}^{n}H}{\mathrm{d}t^n}\rightarrow \infty $, for some $n$ with $n\geq 2$. 
\end{itemize}
In the present work we shall assume that two Type IV singularities occur during the cosmological evolution. From a first glance, the Type IV singularity seems not to affect at all the cosmological system, but as was demonstrated in \cite{noo4}, it can eventually affect the dynamics. As we will demonstrate, the two Type IV singularities offer the possibility of unifying the two accelerating eras of our Universe, with the matter domination era. An intriguing feature of our models is that the deceleration-acceleration transition occurs near the second Type IV singularity.

\section{A Preliminary toy-model: Cosmology Unifying Early and Late-time Acceleration with Matter Domination Eras}

In this section we present in some detail a preliminary cosmological model which describes in a unified way early-time acceleration compatible with observations, late-time acceleration and the matter domination era. In a later section we shall present a variant of this model which describes all the evolution eras of the Universe, but still the qualitative features of both the models are the same. However, we first study the preliminary simplified model, because it is more easy to see the qualitative behavior of the various physical quantities.

The preliminary model has two Type IV singularities as we now demonstrate, with the first occurring at the end of the inflationary era, while the second is assumed to occur at the end of the matter domination era. The chronology of the Universe will assumed to be as follows: The inflationary era is assumed to start at $t\simeq 10^{-35}$sec and is assumed to end at $t\simeq 10^{-15}$sec. After that, the matter domination era occurs, and it is assumed to end at $t\simeq 10^{17}$sec, and after that, the late-time acceleration era occurs. Note that the absence of the radiation era renders the cosmological model just a toy model, but as we mentioned earlier, later on we shall present a variant form of this model which also consistently describes the radiation domination era, in addition to all the other three eras. But the qualitative features of the two models are the same, so we first study this preliminary model for simplicity. So the transition from a decelerated expansion, to an accelerated expansion is assumed to occur nearly at $t\simeq 10^{17}$sec. The Hubble rate of the model is equal to,
\begin{equation}\label{singgener}
H(t)=\e^{-(t-t_s)^{\gamma }} 
\left(\frac{H_0}{2}-H_i (t-t_i)\right)+f_0 |t-t_0|^{\delta } |t-t_s|^{\gamma 
}+\frac{2}{3 \left(\frac{4}{3 H_0}+t\right)}\, ,
\end{equation}
and the values of the freely chosen parameters $t_s$, $H_0$, $t_0$, $\gamma$, $\delta $, $H_i$, $f_0$ and $t_i$, will be determined shortly. For convenience, we shall refer to the cosmological model described by the Hubble rate of Eq. (\ref{singgener}), as the ``unification model''. Before specifying the values of the parameters, it is worth discussing the finite-time singularity structure of the unification model (\ref{singgener}), which will determine the values of the parameters $\gamma$ and $\delta$. Particularly, the singularity structure is the following,
\begin{itemize}\label{lista}
\item When $\gamma,\delta<-1$, then two Type I singularities occurs.
\item When $-1<\gamma,\delta<0$, then two Type III singularities occurs.
\item When $0<\gamma,\delta<1$, then two Type II singularities occurs.
\item When $\gamma,\delta>1$, then two Type IV singularities occurs.
\end{itemize} 
Obviously, there are also more combinations that can be chosen, but we omit these for simplicity. For the purposes of this article, we assume that $\gamma,\delta>1$, so two Type IV singularities occur. Also, as was demonstrated in \cite{noo3,noo4}, if $1<\gamma,\delta<2$, it is possible for the slow-roll indices corresponding to the inflationary era, to develop dynamical instabilities at the singularity points. Also, as was demonstrated in Ref. \cite{noo5}, the gravitational baryogenesis constraints the parameter $\gamma$ to be $\gamma>2$. For these reasons, we assume that $\gamma,\delta>2$. Also, for consistency reasons, we assume that the parameter $\delta$ is of the following form,
\begin{equation}\label{gammaanddelta}
\delta=\frac{2n+1}{2m}\, ,
\end{equation}   
with $n$, and $m$, being positive integers. A convenient choice we shall make for the rest of the paper is that $\gamma=2.1$, $\delta=2.5$. Lets investigate the allowed values of the rest of the parameters, and specifically that of $t_s$, at which the first Type IV singularity occurs. The Type IV singularity at $t=t_s$, will be assumed to occur at the end of the inflationary era, so $t_s$ is chosen to be $t_s\simeq 10^{-15}$sec. Furthermore the second Type IV singularity occurs at $t=t_0$, so at $t_0$ is chosen to be $t_0\simeq 10^{17}$sec. Finally, for reasons to become clear later on, the parameters $f_0$, $H_0$ and $H_i$ are chosen as follows, $H_0\simeq 6.293\times 10^{13}$$\mathrm{sec}^{-1}$, $H_i\simeq 0.16\times 10^{26}$$\mathrm{sec}^{-1}$ and  $f_0=10^{-95}$$\mathrm{sec}^{-\gamma-\delta-1}$. In conclusion, the free parameters in the theory are chosen as follows,
\begin{equation}\label{parameterschoicegeneral}
\gamma=2.1,\,\,\,\delta=2.5,\,\,\,t_0\simeq 10^{17}\mathrm{sec},\,\,\,t_s\simeq 10^{-15}\mathrm{sec},\,\,\,H_0\simeq 6.293\times 10^{13}\mathrm{sec}^{-1},\,\,\,H_i\simeq 6\times 10^{26}\mathrm{sec}^{-1},\,\,\,f_0=10^{-95}\mathrm{sec}^{-\gamma-\delta-1}\, .
\end{equation}
With choice of the parameters as in Eq. (\ref{parameterschoicegeneral}), the model has interesting phenomenology. Firstly let us investigate what happens with the first term of the Hubble rate (\ref{singgener}). Particularly, this term describes the cosmological evolution from $t\simeq 10^{-35}$sec up to $t\simeq 10^{-15}$sec, and it is obvious that the exponential $e^{-(t-t_s)^{\gamma }}$ for so small values of the cosmic time, can be approximated as $e^{-(t-t_s)^{\gamma }}\simeq 1$. In addition, the second term is particularly small during early time, since it contains positive powers of a very small cosmic time and also $f_0$ is chosen to be $f_0=10^{-95}\mathrm{sec}^{-\gamma-\delta-1}$, so the second term can be neglected at early times. Finally, owing to the fact that $t\ll \frac{4}{3 H_0}$, for $10^{-35}<t<10^{-15}$sec, the third term at early times can be approximated as follows,
\begin{equation}\label{earlytimesecondterm}
\frac{2}{3 \left(\frac{4}{3 H_0}+t\right)}\simeq \frac{2}{3 \left(\frac{4}{3 H_0}\right)}=\frac{H_0}{2}\, .
\end{equation}
By combining the above facts, it can be easily seen that the Hubble rate at early times is approximately equal to,
\begin{equation}\label{earlytimehubble}
H(t)\simeq H_0-H_i\left ( t-t_i\right)\, ,
\end{equation}
which is identical to the nearly $R^2$ quasi-de Sitter inflationary evolution \cite{noo4,starobinsky}. This approximate behavior for the Hubble rate at early times holds true for quite a long time after $t\simeq 10^{-15}$sec, and particularly it holds true until the exponential $e^{-(t-t_s)^{\gamma }}$ starts to take values smaller than one, which occurs approximately for $t\simeq 10^{-3}$sec. So for $t>10^{-3}$sec, or more accurately, after $t>1$sec, the exponential term takes very small values, so the first term of the Hubble rate (\ref{singgener}) can be neglected. Then, for a large period of time, the cosmological evolution is dominated by the last term solely, which is,
\begin{equation}\label{matterdomcosmoevolution}
H(t)\simeq \frac{2}{3 \left(\frac{4}{3 H_0}+t\right)}\, ,
\end{equation}
and since $t>1$, and $t\gg \frac{4}{3 H_0}$, for $H_0$ chosen as in Eq. (\ref{parameterschoicegeneral}), the Hubble rate is approximately equal to,
\begin{equation}\label{matterdomcosmoevolution1}
H(t)\simeq \frac{2}{3 t}\, ,
\end{equation}
which exactly describes a matter dominated era, since the corresponding scale factor can be easily shown that it behaves as $a(t)\simeq t^{2/3}$. As we demonstrate shortly, by studying the behavior of the effective equation of state (EoS), we will arrive to the same conclusion. So after the early-time acceleration era, the unification model of Eq. (\ref{singgener}) describes a matter dominated era. This era persists until the present time, with the second term of the Hubble rate (\ref{singgener}) dominating over the last term, only at very late times. So at late-time, the unification model Hubble rate behaves as follows,
\begin{equation}\label{latetimebehavior}
H(t)\simeq f_0 |t-t_0|^{\delta } |t-t_s|^{\gamma 
}\, .
\end{equation}
It is easy to understand why this behavior occurs, since the very small value of the parameter makes the term (\ref{latetimebehavior}) very small, until late times, but then around $t\sim 10^{18}$sec, the term completely overwhelms over the last term of the Hubble rate (\ref{singgener}). In order to further support this result, in Table \ref{table1} we gathered the values of the second and last term of the Hubble rate (\ref{singgener}), for various values of the cosmic time, and also for the values of the parameters chosen as in Eq. (\ref{parameterschoicegeneral}). 
\begin{table*}[h]
\small
\caption{\label{table1}Values of the Various Terms Appearing in the Hubble rate of the Unification Model}
\begin{tabular}{@{}crrrrrrrrrrr@{}}
\tableline
\tableline
\tableline
Term $\,\,\,$ &$\,\,\,\,\,\,\,\,\,\,\,\,\,\,\,$$t=1.2\times 10^{17}$sec & $t=3 \times 10^{17}$sec & $t=10^{18}$sec
\\\tableline
$f_0 |t-t_0|^{\delta } |t-t_s|^{\gamma 
}$ $\,\,\,$ & $4.15\times 10^{-19}$ &$\,\,\,$$9\times 10^{-16}$$\,\,\,$ & $4.8\times 10^{-13}$
\\\tableline
$\frac{2}{3 \left(\frac{4}{3 H_0}+t\right)}$ $\,\,\,$ & $5.55\times 10^{-18}$ &$\,\,\,$$2.2\times 10^{-18}$$\,\,\,$ & $6.6\times 10^{-19}$
\\\tableline
\tableline
\tableline
 \end{tabular}
\end{table*}
As we can see in Table \ref{table1}, indeed the last term of (\ref{singgener}) governs the evolution until very recently (note that the present time is $t=3\times 10^{17}$sec), so the Universe is described by a matter dominated era until very recently and then is described by a late-time acceleration era. In the next section we will make this more clear by studying the behavior of the Hubble radius, in which case, the Type IV singularity that occurs at the end of the matter domination era, seems to affect the deceleration-acceleration transition. 

The same picture we just described can be verified by studying the EoS of the cosmological model of Eq. (\ref{singgener}). Since this model will be described by $F(R)$ gravity models, as was demonstrated in Refs. \cite{reviews1}, the EoS in terms of the Hubble rate reads,
\begin{equation}\label{eosinitialequation}
w_{\mathrm{eff}}=-1-\frac{2\dot{H}(t)}{H(t)^2}\, ,
\end{equation}
so for the Hubble rate of Eq. (\ref{singgener}), the EoS reads,
\begin{align}\label{eosreads}
& w_{\mathrm{eff}}=-1-\frac{2 \left(e^{-(t-t_s)^{\gamma }} H_i-\frac{1}{2 \left(\frac{1}{H_0}+t\right)^2}-e^{-(t-t_s)^{\gamma }} \left(\frac{H_0}{2}+H_i (t-t_i)\right) (t-t_s)^{-1+\gamma } \gamma \right)}{3 \left(\frac{1}{2 \left(\frac{1}{H_0}+t\right)}+e^{-(t-t_s)^{\gamma }} \left(\frac{H_0}{2}+H_i (t-t_i)\right)+f_0 (t-t_0)^{\delta } (t-t_s)^{\gamma }\right)^2}\\ \notag &
-\frac{2 \left(f_0 (t-t_0)^{\delta } (t-t_s)^{-1+\gamma } \gamma +f_0 (t-t_0)^{-1+\delta } (t-t_s)^{\gamma } \delta \right)}{3 \left(\frac{1}{2 \left(\frac{1}{H_0}+t\right)}+e^{-(t-t_s)^{\gamma }} \left(\frac{H_0}{2}+H_i (t-t_i)\right)+f_0 (t-t_0)^{\delta } (t-t_s)^{\gamma }\right)^2}\, .
\end{align}
Therefore, it can be easily shown that at early times, the EoS is approximately equal to,
\begin{equation}\label{eosearlyts}
w_{\mathrm{eff}}\simeq -1-\frac{2 \left(\frac{3 H_0}{4}+H_i\right)}{3 (H_0+H_i 
(t-t_i))^2}\, ,
\end{equation}
so effectively the EoS of this form describes a nearly de Sitter acceleration, since the EoS is very close to $-1$, because the parameters $H_0$ and $H_i$ satisfy $H_0, H_i\gg 1$. After the early times, the EoS can be approximated as follows,
\begin{equation}\label{eosearlyts2}
w_{\mathrm{eff}}\simeq -1-\frac{2 \left(-\frac{2}{3 t^2}\right)}{3 
\left(\frac{2}{3 t}\right)^2}= 0\, ,
\end{equation}
which describes a matter dominated era, since $w_{\mathrm{eff}}\simeq 0$. Note that this behavior is more pronounced as the second Type IV singularity at $t=t_0$ is approached. Finally, at late times, the EoS is approximately equal to,
\begin{equation}\label{eosearlyts4}
w_{\mathrm{eff}}\simeq -1-\frac{2 t^{-1-\gamma -\delta } \gamma }{3 
f_0}-\frac{2 t^{-1-\gamma -\delta } \delta }{3 f_0}\, ,
\end{equation}
which again describes a nearly de Sitter acceleration era, since $f_0$ satisfies $f_0\ll 1$. Note that the EoS (\ref{eosearlyts4}) describes a nearly de Sitter but slightly turned to phantom late-time Universe, a feature which is anticipated and partially predicted for the late-time Universe, see for example \cite{phantom}. But we need to stress that the second and third terms of the EoS in Eq. (\ref{eosearlyts4}), are extremely small, so the difference from the exact de Sitter case can be hardly detected, as time grows. 

To recapitulate, the cosmological model of Eq. (\ref{singgener}) at early times behaves as the $R^2$ inflation model, so it describes an inflationary era, which as we show, it is also compatible with observations. After the early times and until recently in the past, the model (\ref{singgener}) described a matter dominated era and at late times, the model describes a late-time acceleration era. Schematically, the behavior of the model (\ref{singgener}) appears below,
\begin{equation}\label{behaviorofhubble}
H(t)\simeq \left \{ \begin{array}{c}
  H_0-H_i\left ( t-t_i\right),\,\,\,10^{-35}\,\mathrm{sec}<t<10^{-15}\,\mathrm{sec} \\
  \frac{2}{3 t} ,\,\,\,10^{-6}\,\mathrm{sec}<t<10^{17}\,\mathrm{sec} \\
  f_0 |t-t_0|^{\delta } |t-t_s|^{\gamma 
} ,\,\,\,t>2\times 10^{17}\,\mathrm{sec}\\
\end{array}
\right.
\end{equation}
In the next section we thoroughly discuss the behavior of the Hubble radius, or equivalently, of the Hubble horizon. This study will provide extra support to our claims for the behavior of the cosmological model and also will reveal interesting phenomenological implications.

\section{The Hubble Radius Behavior for the Unification Cosmology}

At early times, the cosmological model of Eq. (\ref{singgener}) is described by the quasi de Sitter evolution with Hubble rate appearing in Eq. (\ref{earlytimehubble}), so this is an inflationary evolution. It is worth recalling how the Hubble horizon evolves in the context of a viable cosmological evolution, and in the course we reveal how the cosmological horizon should evolve at late times. As we will demonstrate, the cosmological model has all these interesting features, so it describes a viable cosmological evolution. The inflationary paradigm was introduced in cosmology in order to solve some serious issues of the standard Big Bang theory, such as the initial conditions and flatness issues. But the most important problem which was consistently explained with inflation theory was the following: if the Big Bang theory took place and the Universe evolved from the beginning following only a radiation and matter domination era, then there would exist portions in the Universe, which would be causally disconnected but are observationally very similar. The inflationary theory \cite{inflation} filled the gap, since the exponential evolution at early times could make causally connected eras, to be very much apart, but still being observationally very similar. A very useful quantity which makes the physical description of the cosmological evolution very simple, is the Hubble radius, which is defined to be $R_H=\frac{1}{a(t)H(t)}$, with $a(t)$ being the scale factor and $H(t)$ the corresponding Hubble rate. According to observations, the Hubble radius at the initial singularity was infinite, and decreased very much before and during the inflationary era. After the inflationary era and during the radiation domination era, the Hubble radius started to increase and this behavior continued in the subsequent matter domination era. After the mater domination era, the Universe started to expand but in an accelerating way. This means that the Hubble horizon after the matter domination era, started to decrease again and still decreases even up to today. The era relevant for our cosmological predictions and the very own fact that we are able to make predictions about the early-time cosmological evolution, is due to the fact that at early times, all the cosmologically relevant primordial quantum fluctuation modes of the comoving curvature, were at subhorizon scales, since the comoving wavenumber satisfied $k\gg H(t)a(t)$, or equivalently, the corresponding wavelength satisfied $\lambda \ll \left( H(t)a(t) \right)^{-1}$. As the horizon decreased during inflation, these primordial modes, crossed the horizon when the comoving wavenumber satisfied $k=a(t_H)H(t_H)$, and ``froze'', since these did not evolve in time anymore. As the inflationary era continued in time, the horizon decreased more and more, and the wavenumber of the primordial modes satisfied $k\ll a(t)H(t)$, or the wavelength satisfied $\lambda \gg \left( H(t)a(t) \right)^{-1}$. After the graceful exit from inflation, the Hubble horizon started increasing, so eventually, these primordial modes reentered the horizon, and this is why today we are able to make predictions on the early times cosmology, because we observe practically the power spectrum of these primordial curvature perturbations. Actually, the large scale structure is generated by the gravitational collapse of exactly these primordial modes. So to recapitulate, for a viable cosmological evolution, the Hubble radius initially decreases significantly, from the initial singularity and during the inflationary era, subsequently, during the radiation and matter domination era increases significantly and at the deceleration acceleration transition and after, the Hubble radius started to decrease and still does up to date.

\begin{figure}[h] \centering
\includegraphics[width=16pc]{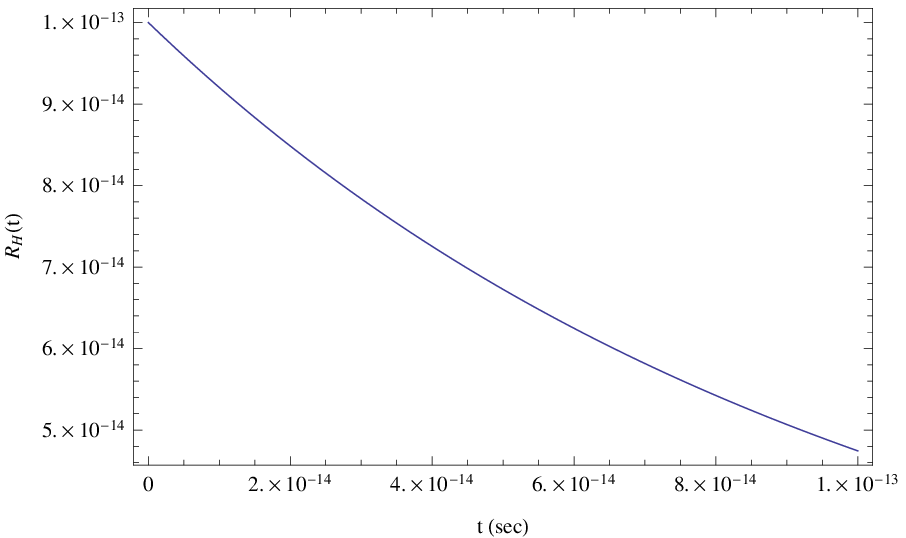}
\includegraphics[width=16pc]{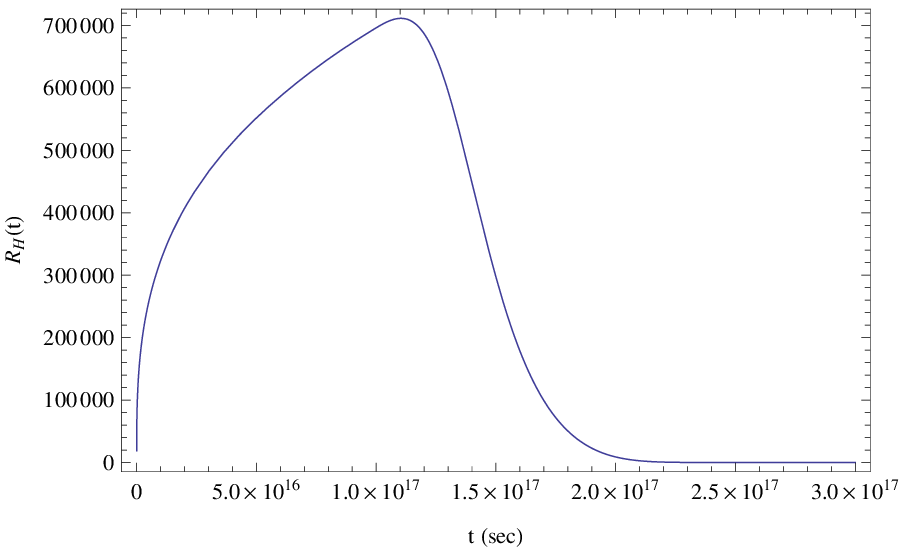}
\caption{The Hubble radius $R_H(t)$ as a function of the cosmic time, for $\gamma=2.1$ $\delta=2.5$, $t_0\simeq 10^{17}\mathrm{sec}$, $t_s\simeq 10^{-15}\mathrm{sec}$, $H_0\simeq 6.293\times 10^{13}\mathrm{sec}^{-1}$, $H_i\simeq 6\times 10^{26}\mathrm{sec}^{-1}$, $f_0=10^{-95}\mathrm{sec}^{-\gamma-\delta-1}$, at early times (left plot) and at later times (right plot).}
\label{plot1}
\end{figure}

As we now demonstrate, in the context of the cosmological model of Eq. (\ref{singgener}), the Hubble horizon initially decreases, then subsequently started to increase and at late times decreases again. But the most intriguing feature of the cosmological evolution of Eq. (\ref{singgener}), is that near the second Type IV singularity, the deceleration to acceleration transition occurs. The rest of this section is devoted to the study of the Hubble horizon evolution of the model (\ref{singgener}). Recall that the Hubble radius is defined to be  $R_H(t)=\frac{1}{a(t)H(t)}$, so we need to compute the scale factor from the Hubble rate (\ref{singgener}). In order to obtain an analytic form of the scale factor, we exploit the fact that at early times, the exponential factor of the first term of Eq. (\ref{singgener}) is nearly equal to one, while at late times it suppresses the first term, so effectively, the scale factor is,
\begin{equation}\label{behaviorofscale}
a(t)\simeq \left \{ \begin{array}{c}
  e^{H_0 t-\frac{H_i t^2}{2}+H_i t t_i},\,\,\,10^{-35}\,\mathrm{sec}<t<10^{-15}\,\mathrm{sec} \\
  e^{\frac{f_0 (t-t_0)^{\delta } (t-t_s)^{1+\gamma } \left(1+\frac{t-t_s}{-t_0+t_s}\right)^{-\delta } \,\,  _2F_1\left(1+\gamma ,-\delta ,2+\gamma ,-\frac{t-t_s}{-t_0+t_s}\right)}{1+\gamma }} (4+3 H_0 t)^{2/3} ,\,\,\,10^{-6}\,\mathrm{sec}<t<10^{17}\,\mathrm{sec} \\
 e^{\frac{f_0 (-t+t_0)^{\delta } (t-t_s)^{1+\gamma } \left(1+\frac{t-t_s}{-t_0+t_s}\right)^{-\delta } \,\,_2 F_1\left(1+\gamma ,-\delta ,2+\gamma ,-\frac{t-t_s}{-t_0+t_s}\right)}{1+\gamma }} (4+3 H_0 t)^{2/3} ,\,\,\,t>\times 10^{17}\,\mathrm{sec}\\
\end{array}
\right.
\end{equation}
where in Eq. (\ref{behaviorofscale}) we took into account the properties of the absolute value of $|t-t_0|^{\delta }$, and also $\, _2F_1(\alpha,\beta,\gamma,x)$ is the Hypergeometric function. The corresponding Hubble radius can easily be found and it is equal to:
 \begin{equation}\label{behaviorofscale1}
R_H(t)\simeq \left \{ \begin{array}{c}
  \frac{e^{-H_0 t+\frac{H_i t^2}{2}-H_i t t_i}}{H_0-H_i t+H_i t_i},\,\,\,10^{-35}\,\mathrm{sec}<t<10^{-15}\,\mathrm{sec} \\
  \frac{e^{-\frac{f_0 (t-t_0)^{\delta } (t-t_s)^{1+\gamma } \left(\frac{-t+t_0}{t_0-t_s}\right)^{-\delta } \,\, _2F_1\left(1+\gamma ,-\delta ,2+\gamma ,\frac{t-t_s}{t_0-t_s}\right)}{1+\gamma }} (4+3 H_0 t)^{1/3}}{H_0 \left(2+3 f_0 t (t-t_0)^{\delta } (t-t_s)^{\gamma }\right)+4 f_0 (t-t_0)^{\delta } (t-t_s)^{\gamma }} ,\,\,\,10^{-6}\,\mathrm{sec}<t<10^{17}\,\mathrm{sec} \\
 \frac{e^{-\frac{f_0 (-t+t_0)^{\delta } (t-t_s)^{1+\gamma } \left(\frac{-t+t_0}{t_0-t_s}\right)^{-\delta } \,\, _2F_1\left(1+\gamma ,-\delta ,2+\gamma ,\frac{t-t_s}{t_0-t_s}\right)}{1+\gamma }} (4+3 H_0 t)^{1/3}}{H_0 \left(2+3 f_0 t (t-t_0)^{\delta } (t-t_s)^{\gamma }\right)+4 f_0 (t-t_0)^{\delta } (t-t_s)^{\gamma }} ,\,\,\,t>\times 10^{17}\,\mathrm{sec}\\
\end{array}
\right.
\end{equation}
In order to have a clear picture on the time dependence of the Hubble radius, or equivalently of the Hubble horizon, in Fig. \ref{plot1} we have plotted the time dependence of the Hubble radius $R_H$, for the values of the parameters chosen as in Eq. (\ref{parameterschoicegeneral}). As it can be seen from the left plot of Fig. \ref{plot1}, the Hubble radius decreases during the inflationary era, as is expected from the form of the scale factor. At later times and during the matter domination era, the Hubble radius increased, while near the second Type IV singularity, which occurs at  $t=t_0$, the Hubble radius starts to decrease again. This behavior is phenomenologically interesting, since the Hubble radius decreases during early-time acceleration, subsequently increases until a time very close to present time, and at $t\sim 10^{17}$sec, where the second Type IV singularity occurs, the Hubble radius starts to decrease again, so the late-time acceleration era starts, and the Universe undergoes late-time acceleration. 
One very interesting feature of the behavior of the Hubble radius, is that the Type IV singularity seems to affect the deceleration-acceleration transition of the Universe. Indeed, during the matter domination era, the Universe expanded in a decelerating way, and near the second Type IV singularity at $t\sim 10^{17}$sec, the Universe started to expand but in an accelerating way, so the deceleration-acceleration occurred near the Type IV singularity at $t=t_0=10^{17}$sec, and the Hubble radius decreased and still decreases until present time. In the next section, we shall investigate how the cosmological evolution of Eq. (\ref{behaviorofhubble}), can be realized by vacuum $F(R)$ gravity.

\section{Realizing the Unification Cosmology with $F(R)$ Gravity}

In the context of $F(R)$ gravity theory, many cosmological scenarios which were ``exotic'' for the standard Einstein-Hilbert gravity, can consistently be realized by using the theoretical framework of $F(R)$ gravity, even in the absence of matter fluids. In this section we investigate which vacuum $F(R)$ gravity can realize the cosmological evolution of Eq. (\ref{singgener}), and to this end, we employ the $F(R)$ gravity reconstruction scheme, which was developed in Refs. \cite{Nojiri:2006gh,Capozziello:2006dj,sergbam08}. Consider the vacuum $F(R)$ gravity action,
\begin{equation}
\label{action1dse}
\mathcal{S}=\frac{1}{2\kappa^2}\int \mathrm{d}^4x\sqrt{-g}F(R)\, .
\end{equation}
By using the metric formalism of $F(R)$ gravity, we vary the Jordan frame action of Eq. (\ref{action1dse}), with respect to the metric tensor $g_{\mu,\nu}$, so the resulting equations of motion are
\begin{equation}
\label{frwf1}
 -18\left ( 4H(t)^2\dot{H}(t)+H(t)\ddot{H}(t)\right )F''(R)+3\left
[H^2(t)+\dot{H}(t)
\right ]F'(R)-
\frac{F(R)}{2}=0\, .
\end{equation}
By using an auxiliary scalar field $\phi$, we rewrite the action of Eq. (\ref{action1dse}), can be cast in the following way,
\begin{equation}
\label{neweqn123}
S=\int \mathrm{d}^4x\sqrt{-g}\left [ P(\phi )R+Q(\phi ) \right ]\, ,
\end{equation}
and since there is no kinetic term for the scalar field $\phi$ in the action (\ref{neweqn123}), the auxiliary scalar field is a non-dynamical field. The aim of the reconstruction technique we employ, is to find the functions $P(\phi (R) )$ and $Q(\phi (R) )$, and in order to find these, upon variation of the action (\ref{neweqn123}) with respect to $\phi$, we obtain,
\begin{equation}
\label{auxiliaryeqns}
P'(\phi )R+Q'(\phi )=0\, ,
\end{equation}
with the prime this time indicating differentiation with respect to $\phi$. We need to stress that, since the actions
of Eqs. (\ref{action1dse}) and
(\ref{neweqn123}) are mathematically equivalent, the auxiliary field and the cosmic time can be identified (for a proof of this claim see the Appendix of Ref.~\cite{Nojiri:2006gh}). By solving the resulting algebraic equation (\ref{auxiliaryeqns}), the function $\phi (R)$ is obtained, and finally by substituting the resulting $\phi (R)$ into Eq.  (\ref{neweqn123}), we obtain the resulting $F(R)$ gravity, which is of the form,
\begin{equation}
\label{r1}
F(\phi( R))= P (\phi (R))R+Q (\phi (R))\, .
\end{equation}
Therefore, the main point of the reconstruction method is to find the functions $P(\phi)$ and $Q(\phi)$. The equations of motion in terms of the functions $P(\phi )$ and $Q(\phi )$, are equal to,
\begin{align}
\label{r2}
 & -6H^2P(\phi (t))-Q(\phi (t) )-6H\frac{\mathrm{d}P\left (\phi (t)\right
)}{\mathrm{d}t}=0\, , \nn
& \left ( 4\dot{H}+6H^2 \right ) P(\phi (t))+Q(\phi (t)
)+2\frac{\mathrm{d}^2P(\phi
(t))}{\mathrm
{d}t^2}+\frac{\mathrm{d}P(\phi (t))}{\mathrm{d}t}=0\, .
\end{align}
So by deleting the function $Q(\phi (t))$ from Eq. (\ref{r2}), we get,
\begin{equation}
\label{r3}
2\frac{\mathrm{d}^2P(\phi (t))}{\mathrm {d}t^2}-2H(t)\frac{\mathrm{d}P(\phi
(t))}{\mathrm{d}t}+4\dot{H}P(\phi (t))=0\, .
\end{equation}
The differential equation (\ref{r3}) will be the starting point for almost all the cases we investigate, so given the Hubble rate $H(t)$, we will solve the differential equation (\ref{r3}) and then we can obtain the function $P(\phi)$, and hence by using Eq. (\ref{r2}), we obtain the function $Q(t)$, and from this we can obtain the function $\phi (R)$ and therefore $F\left(\phi (R)\right )$.

\subsection{Early-time Description-The Nearly $R^2$ Inflation Case}

We start off with the early-time era, in which case, the Hubble rate of the model (\ref{singgener}) is approximately equal to the one appearing in Eq. (\ref{earlytimehubble}). This is the simplest case we shall present since as we now demonstrate, the Hubble rate (\ref{earlytimehubble}) can be generated by an $R^2$ gravity. Indeed, consider the following $F(R)$ gravity,
\begin{equation}\label{starobfr}
F(R)=R+\frac{1}{H_i}R^2\, ,
\end{equation}
which describes the well known $R^2$ inflation model \cite{noo4,starobinsky}. The corresponding FRW equations are equal to,
\begin{equation}\label{takestwo}
\ddot{H}-\frac{\dot{H}^2}{2H}+\frac{M^2}{2}H=-3H\dot{H}\, .
\end{equation}
Owing to the fact that the terms $\ddot{H}$ and $\dot{H}$ can be
neglected during the inflationary era, the Hubble rate corresponding to the $R^2$ model of Eq. (\ref{starobfr}) is equal to,
\begin{equation}\label{hubstar}
H(t)\simeq H_0-H_i\left ( t-t_i\right )\, ,
\end{equation}
which is identical to the Hubble rate (\ref{earlytimehubble}). So at early times, the $F(R)$ gravity that generates the Hubble rate (\ref{earlytimehubble}) is the $R^2$ model of Eq. (\ref{starobfr}). The latest Planck data predict that the $R^2$ inflation model is compatible with the observations, and we now show this explicitly for the Jordan frame model. In the standard $R^2$ inflation model of Eq. (\ref{starobfr}), the inflationary era ends at the moment that one of the Hubble slow-roll parameters
$\epsilon$ becomes of order one, that is, $\epsilon\sim 1$. Recall that the first two Hubble slow-roll parameters are equal to \cite{barrowslowroll} (see also \cite{slowrollindices} for a recent study), 
\begin{equation}\label{hubbleslowrooll}
\epsilon_H=-\frac{\dot{H}}{H^2},\,\,\, \eta_H=-\frac{\ddot{H}}{2H\dot{H}}\,
,
\end{equation}
so in the case at hand, the graceful exit occurs when the perturbative slow-roll expansion breaks at first order. Suppose that the graceful exit occurs at the time instance $t_f$, so the Hubble rate at $t=t_f$ is equal to,
$H_f=H_0-H_i(t_f-t_i)$, and since $\epsilon\simeq 1$ at
$t=t_f$, we obtain that, $H_f\simeq \sqrt{H_i}$. Moreover, from the condition $\epsilon_1(t_f)\simeq 1$, we easily obtain that,
\begin{equation}\label{tfending}
t_f\simeq t_i+\frac{H_0}{H_i}\, .
\end{equation}
We calculate in detail the Hubble slow-roll indices in order to explicitly show that concordance with observations can be achieved. To this end, we shall use the $e$-foldings number $N$, defined as follows,
\begin{equation}\label{efoldings}
N=\int_{t_i}^{t}H(t)\mathrm{d}t\, .
\end{equation}
where for $H(t)$ we will use the expression of Eq. (\ref{earlytimehubble}). In terms of the cosmic time, the Hubble slow-roll indices read,
\begin{equation}\label{hubslowordstarob}
\epsilon_H=\frac{H_i}{6 \left(H_0- H_i (t-t_i)\right)^2},\,\,\,
\eta_H=0\, .
\end{equation}
It is useful to express the Hubble slow-roll parameters in terms of the $e$-foldings number $N$, so by using Eqs. (\ref{efoldings}) and (\ref{hubstar}), we obtain,
\begin{equation}\label{ggdfvgvhsdgd}
t-t_i=\frac{2 \left(3 H_0+\sqrt{3} \sqrt{3 H_0^2-6 H_i N}\right)}{6 H_i}\, ,
\end{equation}
so by substituting into Eq. (\ref{hubslowordstarob}) we finally get,
\begin{equation}\label{epskiolonsokio}
\epsilon_H=\frac{6 H_i}{6 H_0^2-12 H_i N}\, .
\end{equation}
The spectral index of primordial curvature perturbations $n_s$ and the
scalar-to-tensor ratio $r$, expressed in terms of the Hubble slow-roll parameters, for the vacuum $F(R)$ gravity at hand, are defined to be \cite{barrowslowroll},
\begin{equation}\label{egefda}
n_s\simeq 1-4\epsilon_H +2\eta_H,\, \, \,r=48 \epsilon_H^2 \, ,
\end{equation}
which is valid if the parameters remain in the slow-roll regime. Note that since $\eta_H=0$ for the $R^2$ inflation model, the expression for the spectral index of the power spectrum given in Eq. (\ref{egefda}), coincides with the corresponding one given in \cite{slowrollindices}, since, using the Hubble flow parameters $\epsilon_i$ $i=1,..,4$, we have, $\epsilon_1\simeq-\epsilon_3$ and $\epsilon_4\simeq-\epsilon_1$ for the slow-rolling $R^2$ inflation model. By using Eqs. (\ref{epskiolonsokio}) and (\ref{egefda}), the observational indices read,
\begin{equation}\label{gdgfdfdf}
n_s=1-\frac{24\, H_i}{6 H_0^2-12 H_i N},\,\,\, r=48\left(\frac{H_i}{6 H_0^2-12 H_i
N}\right)^2\, .
\end{equation}
The latest Planck data (2015) \cite{planck}, constrain the spectral index and the scalar-to-tensor ratio as follows, 
\begin{equation}\label{recentplancdata}
n_s=0.9655\pm 0.0062\, , \quad r<0.11\, ,
\end{equation}
while the latest BICEP2/Keck-Array data \cite{BICEP2} constrain the scalar-to-tensor ratio as follows,
\begin{equation}\label{keckplancdata}
r<0.07\, .
\end{equation}
Concordance with observations can be achieved if for example, the parameters $H_0$, $H_i$ are chosen to be,
\begin{equation}\label{dkfes}
H_i\sim 6\times 10^{26}\mathrm{sec}^{-1},\,\,\, H_0\sim 6.29348 \times 10^{13}\mathrm{sec}^{-1}\, ,
\end{equation}
and also for $N=60$ $e$-foldings, by substituting these values in Eqs. \ref{gdgfdfdf}, we obtain,
\begin{equation}\label{scealertensorbfer}
n_s\simeq 0.966,\,\,\, r\simeq 0.003468\, ,
\end{equation}
which are compatible with both the latest Planck data of Eq. (\ref{recentplancdata}) and with the BICEP2/Keck-Array data of Eq. (\ref{keckplancdata}). Thus, the cosmological model of Eq. (\ref{singgener}) produces a viable cosmology at early times, compatible with the recent observational data. We need to stress that for the cosmological model (\ref{singgener}), the primordial perturbation modes that are relevant for the present days observations are generated during the inflationary era, when the Hubble radius was decreasing in a nearly exponential rate.

\subsection{$F(R)$ Description of the Matter Domination Era}

The subsequent era of the early-time acceleration one, is the matter domination era, which is described by the third term of the Hubble rate appearing in Eq. (\ref{singgener}). In this section the focus is on finding which $F(R)$ gravity can realize this era. In principle, the optimal case would be, to be able to provide an exact analytic form of the $F(R)$ gravity that can generate the Hubble rate of Eq. (\ref{singgener}), but this is a formidable task, because the resulting differential equations are too difficult to solve analytically. So we split the problem into smaller ones and thus we seek the $F(R)$ gravity that generates the matter domination era solely. In this way we find an approximation of the $F(R)$ gravity corresponding to the matter domination era. Since $H_0\gg 1$ and therefore during the matter domination era, $t\gg \frac{1}{H_0}$, the Hubble rate during the matter domination era is approximately equal to,
\begin{equation}\label{approxhub}
H(t)\simeq \frac{2}{3t}\, ,
\end{equation}
so by using the reconstruction method we presented in the beginning of this section, the resulting differential equation of Eq. (\ref{r3}) becomes,
\begin{equation}
\label{ptdiffeqn}
2\frac{\mathrm{d}^2P(t)}{\mathrm {d}t^2}-\frac{4}{3 t}\frac{\mathrm{d}P(
t)}{\mathrm{d}t}-\frac{8}{3 t^2} P(t)=0\, .
\end{equation}
The differential equation (\ref{ptdiffeqn}) can be solve analytically, and the solution reads,
\begin{equation}
\label{genrealsol}
P(t)= t^{\frac{5}{6}-\frac{\sqrt{73}}{6}} \left(t^{\frac{\sqrt{73}}{3}} C_1+C_2\right)\, ,
\end{equation}
with $C_1$, $C_2$ being arbitrary integration constants. Accordingly, by substituting the resulting expression for $P(t)$ into Eq. (\ref{r2}), we obtain the function $Q(t)$ which reads,
\begin{align}
\label{qtanalyticform}
Q(t) =-6 C_2 t^{-\frac{7}{6}-\frac{\sqrt{73}}{6}}+\frac{2}{3} \sqrt{73} C_2 t^{-\frac{7}{6}-\frac{\sqrt{73}}{6}}-6 C_1 t^{-\frac{7}{6}+\frac{\sqrt{73}}{6}}-\frac{2}{3} \sqrt{73} C_1 t^{-\frac{7}{6}+\frac{\sqrt{73}}{6}}\, .
\end{align}
By using Eqs. (\ref{genrealsol}) and (\ref{qtanalyticform}) and by keeping leading order terms, we substitute the resulting expressions in Eq. (\ref{auxiliaryeqns}), and consequently we obtain,
\begin{equation}
\label{finalxr}
t\simeq \frac{2^{1/6}}{3^{1/12} R^{1/12}} \, .
\end{equation}
Finally, by combining Eqs. (\ref{qtanalyticform}),
(\ref{finalxr}) and (\ref{r1}), the resulting form of the $F(R)$ gravity that generates the matter domination era with Hubble rate (\ref{approxhub}), is equal to,
\begin{equation}
\label{finalfrgravity}
F(R)\simeq (C_1+C_2) R^{\mu }+c_3 R^{\mu +\frac{5}{6}}\, ,
\end{equation}
where the detailed form of the constant parameters $c_i$, $i=1,2,3$ and $\mu$ can be found in the Appendix. Note that in Refs. \cite{Amendola:2006kh,Amendola:2006we}, a power law cosmology described by the first term in Eq. (\ref{finalfrgravity}) is not acceptable, however, we need to note that the result in Refs. \cite{Amendola:2006kh,Amendola:2006we} predicts a full solution for the $F(R)$ gravity of the form $R+\beta/R^{n}$, with $n>0$, whereas in our case, the $F(R)$ gravity of Eq. (\ref{finalfrgravity}) is an approximation of the full $F(R)$ gravity, specified for the corresponding era (something like a Taylor expansion), so this is a leading order description.

\subsection{$F(R)$ Description of the Late-time Era}

In this section we present the approximate form of the $F(R)$ which generates the cosmological evolution of Eq. (\ref{singgener}) at late times. We will be interested in the limit for which $t\gg t_0$, so the Hubble rate can be approximated by,
\begin{equation}\label{approxhub1aa1}
H(t)\simeq f_0 t^{\gamma +\delta }\, ,
\end{equation}
so by utilizing the reconstruction method of the previous sections, the differential equation of Eq. (\ref{r3}) has the following form, 
\begin{equation}
\label{ptdiffeqn1aa1}
2\frac{\mathrm{d}^2P(t)}{\mathrm {d}t^2}-2 f_0 t^{\gamma +\delta }\frac{\mathrm{d}P(
t)}{\mathrm{d}t}4 f_0 t^{-1+\gamma +\delta } (\gamma +\delta ) P(t)=0\, .
\end{equation}
which can be solved analytically, to yield,
\begin{equation}\label{prosolution}
P(t)=2^{\frac{\gamma +\delta }{2 (1+\gamma +\delta )}}C_1 U\left(-\omega_1,\omega_2,\omega_3 t^{1+\gamma +\delta }\right)+2^{\frac{\gamma +\delta }{2 (1+\gamma +\delta )}} C_2 L_{\omega_1}^{1+\omega_2}\left( t^{1+\gamma +\delta }\right)\, ,
\end{equation}
where $U(c,b,x)$ is the confluent hypergeometric function and $L_{n}^{m}(x)$ is the generalized Laguerre polynomial. In addition, the parameters $C_1$ and $C_2$ are arbitrary integration parameters and also the constant parameters $\omega_1$, $\omega_2$ and $\omega_3$ appear in the Appendix.  In order to obtain an analytic solution, we take the large-$t$ limit of the function $P(t)$, so the latter is approximately equal to,
\begin{equation}
\label{genrealsol1aa1}
P(t)= \mathcal{A}_1t^{(1+\gamma +\delta )\omega_1}\, ,
\end{equation}
with $\mathcal{A}_1$ appearing also in the Appendix. By using Eqs. (\ref{genrealsol1aa1}) and (\ref{r2}), the function $Q(t)$ at leading order in the large-$t$ limit reads,
\begin{align}
\label{qtanalyticform1aa1}
Q(t) =-6 \mathcal{A}_1 f_0^2 t^{2 \gamma +2 \delta +(1+\gamma +\delta ) \omega_1}\, .
\end{align}
Finally, by using Eqs. (\ref{genrealsol1aa1}) and (\ref{qtanalyticform1aa1}) and by keeping leading order terms, we obtain,
\begin{equation}
\label{finalxr1aa1}
t\simeq \mathcal{B}_1 R^{\frac{1}{2 \gamma +2 \delta }} \, ,
\end{equation}
with the detailed form of the constant parameter $\mathcal{B}_1$ being given in the Appendix. Correspondingly, the final form of the $F(R)$ gravity that generates the cosmological evolution (\ref{approxhub1aa1}) at late times, is easily found by combining Eqs. (\ref{qtanalyticform1aa1}),
(\ref{finalxr1aa1}) and (\ref{r1}), and it reads,
\begin{equation}
\label{finalfrgravity1aa1}
F(R)\simeq \alpha_1R^{1+\frac{(1+\gamma +\delta ) \omega_1}{2 \gamma +2 \delta }} +\alpha_2R^{\frac{2 \gamma +2 \delta +(1+\gamma +\delta ) \omega_1}{2 \gamma +2 \delta }}\, ,
\end{equation}
which can be further simplified to yield approximately,
\begin{equation}
\label{finalfrgravity1aa11}
F(R)\simeq \left( \alpha_1+\alpha_2 \right )R^{2}\, .
\end{equation}
As a final comment, let us note that the resulting $F(R)$ gravity at very late times is approximately equal to an $R^2$ gravity, but the coefficients $\alpha_1$ and $\alpha_2$, contain powers of $f_0$, which is very small according to our choice of parameters made in Eq. (\ref{parameterschoicegeneral}). Finally, we need to stress that the resulting $F(R)$ realizes the cosmological evolution (\ref{approxhub1aa1}) at cosmic times much more later than the present time, in the limit $t\gg t_0=10^{17}$sec.

\section{A Model Unifying Early and Late-time Acceleration Eras with Radiation and Matter Domination Eras}

Let us here recapitulate our findings up to this point. As we demonstrated in the previous three sections, the vacuum $F(R)$ gravity that can realize the cosmological evolution of Eq. (\ref{singgener}), has the following approximate forms in the three limiting cases, which are at early-time but during the slow-roll inflationary era, during the matter dominating era and during the very late-time acceleration era, later than the present epoch,
\begin{equation}\label{finalfrrecap}
F(R)\simeq \left \{ \begin{array}{c}
  R+\frac{1}{H_i}R^2,\,\,\,10^{-35}\,\mathrm{sec}<t<10^{-15}\,\mathrm{sec} \\
  
  (C_1+C_2) R^{\mu }+c_3 R^{\mu +\frac{5}{6}},\,\,\,10^{-6}\,\mathrm{sec}<t<10^{17}\,\mathrm{sec} \\
 \left( \alpha_1+\alpha_2 \right )R^{2},\,\,\,t> 10^{20}\,\mathrm{sec}\\
\end{array}
\right.
\end{equation}
As we showed in the previous sections, the model (\ref{finalfrrecap}) describes the limiting behavior of the cosmological evolution with Hubble rate (\ref{singgener}). So the model (\ref{finalfrrecap}) is the approximate form of the $F(R)$ gravity which realizes the cosmological evolution (\ref{singgener}), and as we demonstrated this model describes in an unified way, early-time acceleration, late-time acceleration and the matter domination era. Also, the early-time cosmology power spectrum is nearly scale invariant as we demonstrated, and also the predicted scalar-to-tensor ratio is in agreement with the latest observational data. Having in mind these appealing properties, in this section we shall present some variants of the model which also successfully describe a radiation domination era, in addition to the early-late acceleration and matter domination eras. This is a compelling task because the radiation era is a vital feature of a viable and successful cosmological description and should be appropriately described by a cosmological model. 

An interesting generalization of the model (\ref{singgener}) could be the following:
\begin{align}\label{singgenerelazised}
& H(t)=\e^{-(t-t_s)^{\gamma }} 
\left(\frac{H_0}{2}-H_i (t-t_i)\right)+f_0 |t-t_0|^{\delta } |t-t_s|^{\gamma 
}\\ \notag &
+\Theta (t-t_s)\Theta (t_r-t)\frac{1}{2 \left(\frac{1}{H_0}+t\right)}+\Theta (t-t_r)\Theta (t_0-t)\frac{2}{3 \left(\frac{4}{3 H_0}+t\right)}\, ,
\end{align}
with $ \Theta (t)$ being the $\Theta$ step-function, which satisfies,
\begin{equation}\label{finalfrrecap1}
\Theta (t-t_i)\simeq \left \{ \begin{array}{c}
  1,\,\,\,t>t_i\\
  
  0,\,\,\,t<t_i \\
\end{array}
\right.
\end{equation}
The model of Eq. (\ref{singgenerelazised}) describes both the matter and radiation eras and also takes into account for the late and early-time acceleration eras, as the model of Eq. (\ref{singgener}). However, it is somewhat artificial, since the transition from the radiation to the matter domination era is done in a discontinuous way. In addition, the description of the radiation domination era as this appears in the model (\ref{singgenerelazised}), is very hard to be solely described by a vacuum $F(R)$ gravity, since if $a(t)\sim t^{1/3}$, which occurs during a radiation domination era, the corresponding Ricci scalar is $R=0$, when a FRW background is assumed. Therefore, in order to describe the radiation domination era, we need to introduce a relativistic matter fluid accompanying the vacuum $F(R)$ gravity, with EoS parameter $w=1/3$. Hence, it is expected that the relativistic matter fluid with $w=1/3$, will dominate the evolution after the inflationary era, and until the matter domination era takes place. 

Having these issues in mind, in this section we shall present a modification of the unification model of Eq. (\ref{singgener}), with the following characteristics: The model describes in a unified way late and early-time acceleration eras, the radiation and the matter domination eras. The new features of the model is that during the radiation domination era, the EoS of the cosmological system is not exactly $w=1/3$, but it is approximately $w\simeq 1/3$, which means that there is no need to introduce extra matter fluids to account for the radiation domination era, since we can find an approximate form of the $F(R)$ gravity that describes this era, at leading order around the state with $w\simeq 1/3$. To be more quantitative, consider the cosmological model with Hubble rate: 
\begin{equation}\label{newmodel}
H(t)=e^{-(t-t_s)^{\gamma }} 
\left(\frac{H_0}{2}-H_i (t-t_i)\right)+f_0 |t-t_0|^{\delta } |t-t_s|^{\gamma 
}+\frac{1}{\sqrt{3}}\frac{e^{\tanh (t-t_m)\ln \sqrt{\frac{4}{3}}}}{t+\frac{1}{H_0}}\, ,
\end{equation}  
with the time variable $t_m$ characterizing the time that the transition from radiation domination to matter domination occurs, which we assume that it is approximately equal to $t_m\simeq 10^{12}$sec. Comparing the model of Eq. (\ref{singgener}) to the one in Eq. (\ref{newmodel}), we can see that the late and early-time acceleration eras are described by the same terms in the two models, so these are unaffected, and therefore the findings of the previous preliminary cosmological model of Eq. (\ref{singgener}), related to the aforementioned eras, hold true in the model of Eq. (\ref{newmodel}) too. Let us show this explicitly, so by assuming that the parameters take values as chosen in Eq. (\ref{parameterschoicegeneral}), at early times but during the slow-roll inflationary era, for $t<t_s$, the Hubble rate of Eq. (\ref{newmodel}) becomes approximately equal to,
\begin{equation}\label{approxearlynew}
H(t)\simeq H_0-H_i (t-t_i)\, ,
\end{equation}  
since the function $\tanh( t-t_m)$ at early times is approximately equal to,
\begin{equation}\label{tanhearly}
\tanh(t-t_m)\simeq \tanh (-tm)\simeq -1\, ,
\end{equation} 
owing to the fact that $t_m=10^{12}$sec, so it is practically similar to take the limit $\lim_{t\to \infty} \tanh t\simeq -1$. By taking into account the approximation (\ref{tanhearly}), the last term in Eq. (\ref{newmodel}) becomes approximately equal to,
\begin{equation}\label{approlastterm}
\frac{1}{\sqrt{3}}\frac{e^{\tanh (t-t_m)\ln \sqrt{\frac{4}{3}}}}{t+\frac{1}{H_0}}\simeq \frac{1}{\sqrt{3}}\frac{e^{-\ln \sqrt{\frac{4}{3}}}}{t+\frac{1}{H_0}}\simeq \frac{H_0}{2}\, ,
\end{equation}
since $t\ll \frac{1}{H_0}$ during the slow-roll inflation phase. Therefore, at early times, the Hubble rate becomes approximately that of Eq. (\ref{approxearlynew}), which is identical to the one appearing in Eq. (\ref{earlytimehubble}). Therefore, at early times the models (\ref{singgener}) and (\ref{newmodel}) are identical, which means that in this case too, the model (\ref{newmodel}) at early times provides us with a cosmology compatible with the Planck \cite{planck} and the BICEP2/Keck-Array data \cite{BICEP2}.

After the slow-roll inflation era and the graceful exit from inflation, which is assumed to occur at approximately $t\simeq 10^{-15}$sec, the radiation domination era follows until $t\simeq 10^{12}$sec approximately, so let us investigate how the model (\ref{newmodel}) behaves during this era. For the cosmic times in the interval $10^{-12}\mathrm{sec}<t<10^{10}$sec, the function $\tanh (t-t_m)$ is again approximated by (\ref{tanhearly}), so the last term of the Hubble rate (\ref{newmodel}) is approximated by,
\begin{equation}\label{approxlastnew}
\frac{1}{\sqrt{3}}\frac{e^{\tanh (t-t_m)\ln \sqrt{\frac{4}{3}}}}{t+\frac{1}{H_0}}\simeq \frac{1}{\sqrt{3}}\frac{e^{-\ln \sqrt{\frac{4}{3}}}}{t+\frac{1}{H_0}}\simeq \frac{1}{\sqrt{3}}\frac{e^{-\ln \sqrt{\frac{4}{3}}}}{t}\simeq \frac{1}{2t}\, ,
\end{equation}
since in this case, $t\gg \frac{1}{H_0}$. Therefore the Hubble rate of Eq. (\ref{newmodel}), for $10^{-12}\mathrm{sec}<t<10^{10}$sec, is approximately equal to $H(t)\simeq \frac{1}{2t}$, since the first and last terms of Eq. (\ref{newmodel}) are subdominant, as in the case of the model (\ref{singgener}). The Hubble rate $H(t)\simeq \frac{1}{2t}$, generates an EoS which has an EoS parameter approximately equal to $w=\frac{1}{3}$ as we demonstrate soon (see Eq. (\ref{below}) below) by keeping the sub-leading order terms, which means that this describes an approximate radiation era. Hence the model (\ref{newmodel}) for $10^{-12}\mathrm{sec}<t<10^{10}$sec is described by an approximate radiation domination era. Note that the approximation of Eq. (\ref{approxlastnew}) is at leading order, so we shall exploit that later on, by including sub-leading order terms in order to find the $F(R)$ gravity which generates such an approximate cosmological evolution.

We proceed now to the era around $t_m=10^{12}$sec, that is, for cosmic times $10^{11.5}\mathrm{sec}<t<10^{12.5}$sec, in which case the function $\tanh (t-t_m)$, is approximately equal to,
\begin{equation}\label{tanh}
\tanh (t-t_m)\simeq 0\, ,
\end{equation}
and therefore the last term in the Hubble rate of Eq. (\ref{newmodel}) is approximately equal to,
\begin{equation}\label{approxlastnew}
\frac{1}{\sqrt{3}}\frac{e^{\tanh (t-t_m)\ln \sqrt{\frac{4}{3}}}}{t+\frac{1}{H_0}}\simeq \frac{1}{\sqrt{3}t}\, .
\end{equation}
Admittedly, this era is the most peculiar era of the model (\ref{newmodel}), since it corresponds to an EoS which has an EoS parameter which is approximately equal to $w\simeq -1+\frac{2\sqrt{3}}{3}\simeq 0.15$. However, it is obvious that the EoS parameter continuously deforms from $w=1/3$ to a lower value $w\simeq 0.15$, and as we now demonstrate it will reach the approximate value $w\simeq 0$, which describes a matter dominated Universe. Hence, the EoS in the intermediate era $10^{11.5}\mathrm{sec}<t<10^{12.5}$sec is described by a form of collisional matter, as the one studied in \cite{collisional}, so regardless the peculiarity of this short lasting era, the EoS of the Universe is continuously deforming from a radiation dominated EoS to the matter dominated EoS. 

Finally, for the cosmic time in the interval $10^{13}\mathrm{sec}<t<10^{17}$sec, the function $\tanh(t-t_m)$ is approximately equal to,
\begin{equation}\label{tanhlast}
\tanh (t-t_m)\simeq 1\, ,
\end{equation}
so the last term of the Hubble rate (\ref{newmodel}) becomes approximately equal to,
\begin{equation}\label{approxlastnew}
\frac{1}{\sqrt{3}}\frac{e^{\tanh (t-t_m)\ln \sqrt{\frac{4}{3}}}}{t+\frac{1}{H_0}}\simeq \frac{1}{\sqrt{3}}\frac{\ln \sqrt{\frac{4}{3}}}{t+\frac{1}{H_0}}\simeq \frac{2}{3t}\, ,
\end{equation}
and hence the Hubble rate is approximately equal to, $H(t)\simeq \frac{2}{3t}$, which describes a matter dominated Universe with EoS $w\simeq 0$. Finally, after the second Type IV singularity at $t=t_0=10^{17}$sec, the second term of the Hubble rate (\ref{newmodel}) starts to dominate, so the late-time acceleration era starts to occur, and the behavior of the cosmological evolution is identical to the model of Eq. (\ref{singgener}).

\begin{figure}[h] \centering
\includegraphics[width=16pc]{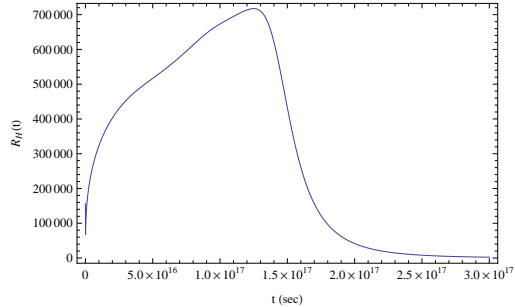}
\caption{The Hubble radius $R_H(t)$ as a function of the cosmic time, for $\gamma=2.1$ $\delta=2.5$, $t_0\simeq 10^{17}\mathrm{sec}$, $t_s\simeq 10^{-15}\mathrm{sec}$, $H_0\simeq 6.293\times 10^{13}\mathrm{sec}^{-1}$, $H_i\simeq 6\times 10^{26}\mathrm{sec}^{-1}$, $f_0=10^{-95}\mathrm{sec}^{-\gamma-\delta-1}$, $t_m=10^{12}$sec.}
\label{hubradiusnew}
\end{figure}

Interestingly enough, the qualitative behavior of the Hubble radius for the model of Eq. (\ref{newmodel}) is identical to the one corresponding to the model (\ref{singgener}) and in Fig. \ref{hubradiusnew} we present the evolution of the Hubble radius as a function of the cosmic time for all eras. We chose the values of the parameters as in Eq. (\ref{parameterschoicegeneral}). As it can be seen, the qualitative behavior is the same, so in this case too, the Hubble horizon after the slow-roll inflation and the graceful exit from inflation, the Hubble horizon expands during the radiation domination era and also during the matter domination eras, until the time scale of the second Type IV singularity, near $t=10^{17}$sec, at the vicinity of which the deceleration acceleration transition occurs, and the Universe starts to accelerate until the present time. 
Therefore, the resulting qualitative behavior for the two models is the same, but in the case at hand, the radiation domination era is described consistently. Particularly, some appealing features of the model (\ref{newmodel}) are:
\begin{itemize}
    \item The late and early-time acceleration are described in a unified way with the matter and radiation domination eras. So all the known eras of the Universe are described consistently.
    \item The transition from the radiation era to the matter domination era is done in a continuous way and also the corresponding EoS is continuously deformed from $w\simeq 1/3$ to $w\simeq 0$. 
    \item The Hubble radius decreases during the slow-roll inflationary era, the increases during the radiation and matter eras and starts to decrease at late times, thus it behaves as the Hubble horizon of a viable cosmology behaves.
\end{itemize}    

Finally, for the model of Eq. (\ref{newmodel}), it is possible to find which $F(R)$ gravity can generate the nearly matter dominated era, since the Hubble rate of radiation domination era is approximately, but not exactly equal to $\sim \frac{1}{3t}$. Hence we can find the approximate Hubble rate at leading order and after that by calculating the Ricci scalar, it can be seen that it is not equal to zero, but very close to zero at leading order. Therefore, the $F(R)$ gravity description of the nearly radiation domination era can be given, at least at leading order. Recall that it was impossible for a vacuum $F(R)$ gravity to describe the radiation domination era, without including a matter fluid, but in the case at hand, we can find a leading order $F(R)$ gravity description. Let us explicitly demonstrate how to find the $F(R)$ gravity at leading order and we start off by finding a leading order approximation of the Hubble rate (\ref{newmodel}), which during the radiation domination era, it can be approximated by the following expression,
\begin{equation}\label{approxhub}
H(t)\simeq \frac{1}{2t}+\frac{\mathcal{D}_1}{t}\, ,
\end{equation}
with the parameter $\mathcal{D}_1$ being equal to,
\begin{equation}\label{d1}
\mathcal{D}_1=-2^{-1} (2 \ln(2)-\ln(3)) \left(-1+\tanh(t_m)^2\right)\, ,
\end{equation}
which since $t_m=10^{12}$sec, satisfies $\mathcal{D}_1\ll 1$. It is easy to find the approximate expression for the EoS parameter $w_{\mathrm{eff}}$ and by combining Eqs. (\ref{eosinitialequation}) and (\ref{approxhub}), the EoS parameter reads,
\begin{equation}\label{below}
w_{\mathrm{eff}}\simeq \frac{1}{3}-\frac{8 \mathcal{D}_1}{3}+\frac{16 \mathcal{D}_1^2}{3}\, ,
\end{equation}
where we expanded in terms of the parameter $\mathcal{D}_1$. As it can be seen from Eq. (\ref{below}) the EoS parameter $w_{\mathrm{eff}}$ is approximately equal to $w\simeq 1/3$, since $\mathcal{D}_1\ll 1$.

Getting back to the vacuum $F(R)$ description of the radiation era, as in the other cases, we shall use the reconstruction method of the previous section, so the differential equation of Eq. (\ref{r3}) for the Hubble rate chosen as in Eq. (\ref{approxhub}), becomes equal to, 
\begin{equation}
\label{ptdiffeqn1aa1radiation}
4 \left(-\frac{1}{2 t^2}-\frac{\mathcal{D}_1}{t^2}\right) P(t)-2 \left(\frac{1}{2 t}+\frac{\mathcal{D}_1}{t}\right) P'(t)+2 P''(t)=0\, .
\end{equation}
and the solution can be found analytically, and it is equal to,
\begin{equation}\label{prosolutionradiation}
P(t)=t^{\frac{3+8 \mathcal{D}_1+4 \mathcal{D}_1^2+\sqrt{1+2 \mathcal{D}_1} \sqrt{25+94 \mathcal{D}_1+92 \mathcal{D}_1^2+8 \mathcal{D}_1^3}}{4+8 \mathcal{D}_1}} C_1+t^{\frac{3+8 \mathcal{D}_1+4 \mathcal{D}_1^2-\sqrt{1+2 \mathcal{D}_1} \sqrt{25+94 \mathcal{D}_1+92 \mathcal{D}_1^2+8 \mathcal{D}_1^3}}{4+8 \mathcal{D}_1}} C_2\, ,
\end{equation}
where $C_1$ and $C_2$ are arbitrary integration parameters. At order zero in the parameter $\mathcal{D}_1$, the function $P(t)$ reads,
\begin{equation}\label{simplify}
P(t)\simeq \frac{C_2}{\sqrt{t}}+C_1 t^2\, ,
\end{equation}
and we shall use the expression (\ref{simplify}) in order to simplify the resulting expressions and in order to be able to treat analytically the problem. However, the parameter $\mathcal{D}_1$ will appear to the function $Q(t)$ due to the presence of the Hubble rate in Eq. (\ref{r2}), but the resulting expression is much more easy to handle analytically and it reads,
\begin{align}
\label{qtanalyticform1aa1radiation}
Q(t) =-\frac{15 C_2}{2}-18 C_2 \mathcal{D}_1-6 C_2 \mathcal{D}_1^2-\frac{3 C_1 \mathcal{D}_1}{t^{5/2}}-\frac{6 C_1 \mathcal{D}_1^2}{t^{5/2}}\, .
\end{align}
Plugging in Eqs. (\ref{simplify}) and (\ref{qtanalyticform1aa1radiation}) in Eq. (\ref{auxiliaryeqns}), by solving with respect to the cosmic time $t$, we obtain,
\begin{equation}
\label{finalxr1aa1radiation}
t\simeq \frac{15^{2/9} C_1^{2/9} \mathcal{D}_1^{2/9} (1+2 \mathcal{D}_1)^{2/9}}{2^{4/9} C_2^{2/9} R^{2/9}} \, ,
\end{equation}
Finally, by combining Eqs. (\ref{qtanalyticform1aa1radiation}),
(\ref{finalxr1aa1radiation}) and (\ref{r1}), we obtain the approximate form of the vacuum $F(R)$ gravity which generates the nearly radiation domination Hubble rate of Eq. (\ref{approxhub}), which is,
\begin{align}
\label{finalfrgravity1aa1radiation}
& F(R)\simeq -\frac{15 C_2}{2}-18 C_2 \mathcal{D}_1-6 C_2 \mathcal{D}_1^2+\frac{15^{4/9} C_1^{4/9} C_2^{5/9} \mathcal{D}_1^{4/9} (1+2 \mathcal{D}_1)^{4/9} R^{5/9}}{2^{8/9}}\\ \notag &
-\frac{2\ 2^{1/9} 3^{4/9} C_1^{1/3} C_2^{2/3} \mathcal{D}_1^{1/3} \sqrt{\frac{C_1^{2/9} \mathcal{D}_1^{2/9} (1+2 \mathcal{D}_1)^{2/9}}{C_2^{2/9} R^{2/9}}} R^{2/3}}{5^{5/9} (1+2 \mathcal{D}_1)^{2/3}}-\frac{4\ 2^{1/9} 3^{4/9} C_1^{1/3} C_2^{2/3} \mathcal{D}_1^{4/3} \sqrt{\frac{C_1^{2/9} \mathcal{D}_1^{2/9} (1+2 \mathcal{D}_1)^{2/9}}{C_2^{2/9} R^{2/9}}} R^{2/3}}{5^{5/9} (1+2 \mathcal{D}_1)^{2/3}}\\ \notag &
+\frac{2^{2/9} C_1^{7/9} C_2^{2/9} \sqrt{\frac{C_1^{2/9} \mathcal{D}_1^{2/9} (1+2 \mathcal{D}_1)^{2/9}}{C_2^{2/9} R^{2/9}}} R^{11/9}}{15^{1/9} \mathcal{D}_1^{2/9} (1+2 \mathcal{D}_1)^{2/9}}
\, ,
\end{align}
and it can be further simplified by taking into account that the approximate expression for the Ricci scalar of Eq. (\ref{ricciscal}) corresponding to the Hubble rate (\ref{approxhub}) is equal to,
\begin{equation}\label{riccscala}
R=\frac{6 \mathcal{D}_1}{t^2}+\frac{12 \mathcal{D}_1^2}{t^2}\, ,
\end{equation}
which means that for large $t$, the Ricci scalar is small. Hence, for most of the time that the radiation domination era lasts, the Ricci scalar is small, so the $F(R)$ gravity of Eq. (\ref{finalfrgravity1aa1radiation}) is approximately equal to,
\begin{equation}
\label{finalfrgravity1aa11radiation}
F(R)\simeq -\frac{15 C_2}{2}-18 C_2 \mathcal{D}_1-6 C_2 \mathcal{D}_1^2+\frac{15^{4/9} C_1^{4/9} C_2^{5/9} \mathcal{D}_1^{4/9} (1+2 \mathcal{D}_1)^{4/9} R^{5/9}}{2^{8/9}}\, .
\end{equation}
Note that the presence of the parameter $\mathcal{D}_1$ in the resulting expressions of the vacuum $F(R)$ gravity which generates the nearly radiation domination era of Eq. (\ref{approxhub}), indicates that this is a leading order result and since $\mathcal{D}_1\ll 1$, the contribution is very small.  Lastly, with regards to the other three evolution eras, the leading order $F(R)$ gravities are the ones appearing in Eq. (\ref{finalfrrecap}), so the sub-leading order terms do not alter the final picture of the $F(R)$ gravity description. In all cases, it is possible to include matter fluids which will alter the final functional form of the $F(R)$ gravity, but we omit this study since this can easily be done and only the functional form of the resulting $F(R)$ gravity will change, the evolution picture will be qualitatively the same.

\section{Discussion and Conclusions}

In this paper we presented two cosmological models which unify the late and early-time acceleration eras with the matter domination era and radiation era, and we provided a consistent realization of these cosmologies by using the theoretical framework of vacuum $F(R)$ modified gravity. A vital feature for the viability of the cosmological evolutions we presented is the appearance of two Type IV singularities, which occur at the end of the inflationary era and at the end of the matter domination era respectively. The cosmological models at early times have the appealing feature of being approximately identical to the $R^2$ inflation model, so this produces a nearly scale invariant power spectrum of primordial curvature perturbations, which is compatible with the latest Planck data \cite{planck}. In addition, the predicted scalar-to-tensor ration is compatible to both the latest Planck and BICEP2/Keck-Array data, so this makes the early-time behavior of the models quite appealing. The graceful exit from inflation in the cosmological models we presented occurs when the slow-roll perturbative expansion breaks down, and this occurs when the first order Hubble slow-roll parameters become of the order one. After that, the radiation and matter domination era occur and last until the second Type IV singularity occurs, which is chosen to occur at $t\sim 10^{17}$sec, quite near the present time epoch. Near the second Type IV singularity and particularly, right after, the matter domination era stops and the Universe starts to expand in an accelerating way, with an equation of state slightly turned to phantom, as we evinced. This kind of behavior is expected for the late-time Universe \cite{phantom}, so this is an interesting feature of the cosmological evolution under study. By utilizing well known reconstruction techniques for modified gravity, we investigated how the cosmological models under study, can be realized in the context of vacuum $F(R)$ gravity. Due to the lack of analyticity, we found the approximate forms of the $F(R)$ gravity for the late-early-time acceleration eras and for the matter and radiation domination eras separately.

Another important study we performed is related to the evolution of the Hubble horizon, as a function of the cosmic time. In this case, we focused our study from the time that the slow-roll inflationary regime started, until the present time epoch and slightly later. The early-time behavior verified our expectations that the Hubble horizon shrinks during the slow-roll inflation, until the radiation and matter domination eras start. After that, and as the Universe expands in a decelerating way, the Hubble horizon starts to expand again. One novel interesting feature of our cosmological model is that the Hubble horizon expands during the radiation and matter domination era, and it stops expanding near the second Type IV singularity which occurs at the end of the matter domination era. After the second Type IV singularity, the Hubble horizon starts to shrink again, since the Universe expands in an accelerating way. The notable feature is that it seems that the second Type IV singularity affects the deceleration acceleration transition, but our study revealed only an indication on this, however no rigid proof. This issue should be further scrutinized, because it can be an artifact of the choice of the parameters, and we intend to undertake this task in a future work focused on exactly this deceleration-acceleration issue. 

Another interesting issue would be to investigate the pre-inflationary era, in order to provide a more complete description of the model. Particularly, there exist various proposals in the literature in which the inflationary era is subsequent to a superinflationary phase \cite{piao1} or to a bouncing phase. The important feature of both the theoretical proposals we just mentioned, is that the initial singularity is avoided. In the models we worked out in this paper, we were not interested at all for this preinflationary era, since we focused on the slow-roll inflationary regime and after. But there is strong motivation to look for a preinflationary era, since the cosmic microwave background has certain features which are not in concordance with a slow-roll inflationary era and particularly, the large scale power deficit cosmic microwave background TT-mode. This issue has been verified by the Planck collaboration \cite{planck} and has been pointed out in the literature \cite{piao1}. As was claimed in \cite{piao1}, the large scale anomalies can be attributed to the physics before the slow-roll era, which can be described by a contracting or expanding bouncing phase or a superinflationary phase. Then, the power spectrum of primordial curvature perturbations receives a large-scale cutoff which can naturally explain the power deficit in the cosmic microwave background TT-mode. The possibility of having a bouncing phase preceding the slow-roll inflationary phase is particularly interesting, since in this way the initial singularity problem can be resolved, and this type of scenario has been considered in the literature \cite{piao1,piao2,piao3,piao4}, and it is called ``the bounce inflation scenario''. A particularly interesting possibility is to have a bounce with a Type IV singularity at the bouncing point \cite{noo3}. As it was demonstrated in \cite{noo3}, the singular bounce in the context of $F(R)$ modified gravity produces a non-scale invariant power spectrum, so it is certain that this scenario alone cannot describe a viable cosmological evolution compatible with the observational data. However, it is possible that the singular bounce describes the era before the slow-roll inflation, which is very likely since the Hubble horizon at the singular point is infinite and drops after the singular point. For a detailed analysis of the Hubble horizon in the context of the singular bounce, see the recent study \cite{nojirisergeinewunibounce} and for other related bouncing cosmologies, see \cite{harobounce}. The modified gravity framework provides the theoretical tools to realize bouncing cosmologies without the need for violation of the null energy conditions, so a realization of these cosmological scenarios in the context of $F(R)$ gravity seems quite appealing. We hope to address some of these issues in a detailed work in the future. For a recent relevant work on preinflationary issues, see \cite{wette}.

Before closing we need to mention that the cosmological scenarios we studied in this paper, can be investigated in the context of Loop Quantum Cosmology (LQC) \cite{LQC}. Particularly interesting would be to find the effects of loop quantum corrections in the equations of motion, and also to investigate if the cosmological scenario can be harbored in the theory of LQC. Note that the issue of LQC Jordan frame quantization of $F(R)$ theories has been analyzed in Refs. \cite{Zhang:2011vi,Zhang:2011qq}. Lastly, note that the inclusion of usual perfect matter fluids can easily be incorporated in the models we presented, but this is not expected to change the overall qualitative behavior of the cosmological evolution of the models, but would definitely change the functional form of the resulting $F(R)$ gravity.

\section*{Acknowledgments}

This work is supported by MINECO (Spain), project
 FIS2013-44881 (S.D.O) and by Min. of Education and Science of Russia (S.D.O
and V.K.O).

\section*{Appendix: Detailed Form of the Parameters Appearing in the Main
Text}

In this Appendix we present the detailed form of various parameters appearing in the text. We start off with the parameters $c_i$, $i=1,2,3$ and $\mu$ appearing in Eq. (\ref{finalfrgravity}), the detailed form of which is,
\begin{align}\label{detailedformofparameters}
& c_1=-2^{\frac{29}{36}+\frac{\sqrt{73}}{36}} 3^{\frac{79}{72}-\frac{\sqrt{73}}{72}} C_1,\,\,\,c_2=-2^{\frac{29}{36}+\frac{\sqrt{73}}{36}} 3^{-\frac{65}{72}-\frac{\sqrt{73}}{72}} \sqrt{73} C_1\\ \notag &
c_3=2^{\frac{5}{36}+\frac{\sqrt{73}}{36}} 3^{-\frac{5}{72}-\frac{\sqrt{73}}{72}} C_1,\,\,\, \mu=\frac{7}{72}-\frac{\sqrt{73}}{72}
\end{align}
Moreover, the parameters $\omega_1$, $\omega_2$ and $\omega_3$ appearing in Eq. (\ref{prosolution}) are equal to,
\begin{equation}\label{parametersomega}
\omega_1=\frac{2 (\gamma +\delta )}{1+\gamma +\delta },\,\,\,\omega_2=\frac{\gamma +\delta }{1+\gamma +\delta },\,\,\, \omega_3=\frac{f_0 }{1+\gamma +\delta }\, .
\end{equation}
Also the parameter $\mathcal{A}_1$ appearing in Eq. (\ref{genrealsol1aa1}), is equal to,
\begin{equation}\label{mathcala1}
\mathcal{A}_1=\frac{2^{\frac{\gamma +\delta }{2 (1+\gamma +\delta )}} \omega_3^{\omega_1} (C_2+C_1 \Gamma (1+\omega_1))}{\Gamma (1+\omega_1)}\, .
\end{equation}
In addition, the parameter $\mathcal{B}_1$ appearing in Eq. (\ref{finalxr1aa1}) is equal to,
\begin{equation}\label{b1forfinalequation}
\mathcal{B}_1=6^{-\frac{1}{2 (\gamma +\delta )}} (\mathcal{A}_1 (1+\gamma +\delta ) \omega_1)^{\frac{1}{2 \gamma +2 \delta }} \left(-\mathcal{A}_1 f_0^2 (\omega_1+\gamma  (2+\omega_1)+\delta  (2+\omega_1))\right)^{-\frac{1}{2 (\gamma +\delta )}}\, .
\end{equation}
Finally, the parameters $\alpha_1$ and $\alpha_2$ appearing in Eq. (\ref{finalfrgravity1aa1}) are equal to,
\begin{equation}\label{finalparameters}
\alpha_1=\mathcal{A}_1 \mathcal{B}_1^{(1+\gamma +\delta ) \omega_1},\,\,\, \alpha_2=-6 \mathcal{A}_1 f_0^2 \mathcal{B}_1^{2 \gamma +2 \delta +(1+\gamma +\delta ) \omega_1}
\end{equation}

\end{document}